\documentclass[12pt]{article} 
\usepackage{amsmath}	
\usepackage{amssymb}	
\usepackage{amsthm}
\usepackage[english]{babel}
\usepackage[latin1]{inputenc} 
\usepackage[T1]{fontenc} 
\usepackage{titlesec}	
\titleformat{\section}{\normalsize\bfseries}{\thesection}{1em}{}
\titleformat{\subsection}{\normalsize}{\thesubsection}{1em}{}

\begin{document}
\title{Optimal partial hedging in a discrete-time market as a knapsack problem}
\author{Peter Lindberg\\ \scriptsize{Department of Mathematics, Åbo Akademi University, Fänriksgatan 3, 20500 Åbo, Finland}\\ \scriptsize{email: plindber@abo.fi}}

\date{}


\newtheorem{theorem}{Theorem}[section]
\newtheorem{proposition}[theorem]{Proposition}
\newtheorem{corollary}[theorem]{Corollary}
\newtheorem{lemma}[theorem]{Lemma}
\newtheorem{definition}[theorem]{Definition}
\newtheorem{remark}[theorem]{Remark}
\newtheorem{example}[theorem]{Example}
\newtheorem*{problema}{Problem A}
\newtheorem*{problemb}{Problem B}
\newtheorem*{problemc}{Problem C}
\newtheorem*{problemd}{Problem D}
\newtheorem*{problemap}{Problem $\bf A'$}
\newtheorem*{problemcp}{Problem $\bf C'$}
\newtheorem*{problemdp}{Problem $\bf D'$}

\numberwithin{table}{section}

\maketitle

\abstract
We present a new approach for studying the problem of optimal hedging of a European option in a finite and complete discrete-time market model. We consider partial hedging strategies that maximize the success probability or minimize the expected shortfall under a cost constraint and show that these problems can be treated as so called \emph{knapsack problems}, which are a widely researched subject in linear programming. This observation gives us better understanding of the problem of optimal hedging in discrete time.

\section{Introduction}

In this paper we work with a finite, arbitrage-free and complete market model in discrete time. It is a well-known fact that a European option can be hedged perfectly in such a market and that the price of a perfect hedge is equal to the unique arbitrage-free price of the option. However, the seller of the option might not be willing to use all the money he has received from selling the option to construct a hedging strategy. She/he may instead want to create only a partial hedge. This kind of a hedge costs less than a perfect one but forces the investor to face the risk of a shortfall.

There are different ways to measure the risk that an investor who uses a partial hedge must take. In this paper the optimality is measured in terms of success probability (i.e. the probability that shortfall will not occur) and in terms of expected shortfall, when a cost constraint is given. Föllmer and Leukert \cite{Föllmer99}, \cite{Föllmer00} and Föllmer and Schied \cite{Föllmer04} have studied these problems in both discrete and continuous time. They search an optimal solution among admissible strategies, i.e. strategies which are self-financing and whose value processes are non-negative. Their solution techniques are mainly based on different applications of Ney\-man-Pearson lemma. Runggaldier, Trivellato and Vargiolu \cite{Runggaldier}, Favero \cite{Favero01}, Favero and Vargiolu \cite{Favero06} and Scagnellato and Vargiolu \cite{Scagnellato} study the problems via dynamic programming in binomial and multinomial models in the case where the strategies are only required to be self-financing.

The main contribution of this paper is that these types of partial hedging problems can be reduced to knapsack problems (see Section \ref{nqp} for a description of a knapsack problem). In particular, this new approach allows us to prove some of the existing results in an alternative way. Gathering the results under knapsack theory helps us also to see interestingly how admissibility condition affects the optimal solutions.

The partial hedging problems covered in this paper are presented in Section \ref{g3}. In Section \ref{doe}, it is shown that the problem of maximizing success probability under admissibility constraint can be reduced to a 0-1 knapsack problem. An approximative algorithm for solving the problem is obtained by studying a related continuous knapsack problem. Comparison with the Neyman-Pearson-based results in Föllmer and Schied \cite{Föllmer04} is also carried out. In Section \ref{lkjh} we will prove that if admissibility constraint is omitted, the success probability is maximized by an optimal \emph{quasi-replicating} strategy, i.e. a strategy that replicates the option in all but one state. In Sections \ref{deo} and \ref{hjkl} we consider the problem of finding a strategy that minimizes the expected shortfall. A knapsack problem approach is used to show how admissibility constraint affects the solutions. An alternative proof for the result in Scagnellato and Vargiolu \cite{Scagnellato}, p.148, will be given.

\section{Partial hedging problems and knapsack problems} \label{g3}

\subsection{Market model}
The market model that we work with in this paper is based on a finite filtered probability space $(\Omega, \mathcal{F},\{\mathcal{F}_i\}_{i=0}^T, \mathbb{P}),$ where $\Omega=\{\omega_1,\ldots,\omega_n\},$ $\mathcal{F}_0=\{\emptyset,\Omega\}$, $\mathcal{F}_T=\mathcal{F}=\mathcal{P}(\Omega)$ and $\mathbb{P}(\{\omega_i\})>0 \mbox{ for all } \omega_i \in \Omega.$ Here $\mathcal{P}(\Omega)$ denotes the set of all subsets of $\Omega$. We set $T$ to be equal to the maturity date of the European option which we want to hedge. 

The prices of the $d+1$ assets on the market follow a $d+1$-dimen\-sional non-negative $\{\mathcal{F}_i\}$-adapted stochastic process $\{\overline{S}_t, t=0,\ldots,T\}$, where $$\overline{S}_t=(S_t^{(0)},S_t^{(1)}, \ldots, S_t^{(d)})$$ and $S_t^{(i)}$ is the value of asset $i$ at time $t$. However, we will follow the approach in Föllmer and Schied \cite{Föllmer04} and present all values in units of the \emph{numéraire} asset $S^0$, whose value therefore is assumed to be strictly positive at all times. The discounted price of asset $i$ at time $t$ (i.e. its value in units of the numéraire asset $S^0$) is given by $X_t^{(i)}:=S_t^{(i)}/S_t^{(0)}$ and the corresponding price process is $\{\overline{X}_t,t=0,\ldots,T\}$, where $$\overline{X}_t=(1,X_t^{(1)}, \ldots, X_t^{(d)}).$$

A trading strategy is an $\mathbb{R}^{d+1}$-valued $\{\mathcal{F}_i\}$-predictable stochastic process $\overline{\xi}=\{\overline{\xi}_t,t=0,\ldots,T\}$, where $$\overline{\xi}_t=(\xi_t^{(0)},\xi_t^{(1)}, \ldots, \xi_t^{(d)})$$ and $\xi_t^{(i)}$ is the quantity of asset $i$ in the portfolio at time $t$. The value process $V=\{V_t,t=0,\ldots,T\}$ of $\overline{\xi}$ is defined through $$V_t= \overline{\xi}_t \cdot \overline{X}_t=\sum_{i=0}^d \xi_t^{(i)}X_t^{(i)}.$$
A strategy $\overline{\xi}$ is called self-financing if $$\overline{\xi_t} \cdot \overline{X}_t = \overline{\xi}_{t+1} \cdot \overline{X}_t$$ for all $t=0,\ldots,T-1$. A self-financing strategy $\overline{\xi}$ is called admissible if its value process satisfies $V_t \ge 0$ for all $t=0,\ldots,T$.

We assume that the market is arbitrage-free and complete and let $\mathbb{P}^*$ denote the unique equivalent martingale measure. Moreover, $H$ is a discounted European contingent claim, i.e. a non-negative random variable on $(\Omega,\mathcal{F}_T,\mathbb{P})$. It is known that $H$ can be hedged perfectly, i.e. there exists a self-financing strategy $\overline{\xi}$ such that $H=\overline{\xi}_T \cdot \overline{X}_T.$ The initial cost of this replicating strategy is equal to the unique discounted arbitrage-free price of $H$ and is given by  $\pi^H=\mathbb{E}^*(H),$ where $\pi^H$ is expressed in units of the numéraire asset.
\begin{remark} \label{comp1}
Completeness is usually defined to mean that every non-negative $\mathcal{F}_T$-meas\-urable random variable can be replicated. However, it is easy to show that in our finite market the same holds for an arbitrary random variable. This can be done by studying separately its positive and negative part.
\end{remark}
\begin{remark}
It is well known that the value process $V$ of a self-financing strategy $\overline{\xi}$ satisfies $$V_t=V_0+\sum_{i=1}^t \overline{\xi}_i \cdot (\overline{X}_i-\overline{X}_{i-1}).$$ Moreover, in a finite probability space it is easy to show that $V$ is always a $\mathbb{P}^*$-martingale.
\end{remark}

\subsection{Partial hedging problems}
Next we fix an upper bound $v<\mathbb{E}^*(H)$ for the initial payment (in numéraire units) that the investor wants to use for creating a hedging strategy. Under this cost constraint we try to find an optimal partial hedging strategy. The problems considered in this paper are stated below. In problems A and B optimality is measured in terms of success probability, in C and D through expected shortfall. Note that in problems A and C the optimal solution is searched among admissible strategies, whereas in B and D we only require that the strategies are self-financing.

\begin{problema}
\noindent Find an admissible strategy whose value process $V$ maximizes $\mathbb{P}(V_T \ge H)$ under the constraint $V_0 \le v.$
\end{problema}
\begin{problemb}
Find a self-financing strategy whose value process $V$ maximizes $\mathbb{P}(V_T \ge H)$ under the constraint $V_0 \le v.$
\end{problemb}
Referring to the discussion in Föllmer and Schied \cite{Föllmer04}, p.341 we state that the problem of minimizing the expected shortfall $\mathbb{E}[(H-V_T)^+]$ can be simplified to the following problems:
\begin{problemc}
Find an admissible strategy whose value process $V$ maximizes $\mathbb{E}(V_T)$ under constraints $V_T \le H$ and $V_0 \le \nolinebreak v$.
\end{problemc}
\begin{problemd}
Find a self-financing strategy whose value process $V$ maximizes $\mathbb{E}(V_T)$ under constraints $V_T \le H$ and $V_0 \le \nolinebreak v$.
\end{problemd}
\subsection{The knapsack problem} \label{nqp}

The main purpose of this paper is to show that the partial hedging problems A, C and D above can be reduced to knapsack problems. The knapsack problem is usually illustrated as follows (see e.g Dantzig \cite{Dantzig57}, p.273 or Martello and Toth \cite{Martello}, p.1): A traveller has to fill a knapsack of a certain size $c$ by selecting some of $n$ objects having sizes $w_i$, $i=1,\ldots,n$, respectively. The ``comfort'' or ``gain'' given by the objects is measured with numbers $g_i$, $i=1,\ldots,n$, respectively. The traveller wants to select objects that give her/him the maximal total ``comfort'' or ``gain'' under the constraint that the total size of the chosen objects will not exceed the knapsack size $c$. We model a possible decision by an $n$-dimensional binary vector $x$ whose elements satisfy  $$x_i=\left\lbrace 
\begin{array}{ll}
1 & \rm{if} \mbox{ } \rm{object}\mbox{ } \emph{i} \mbox{ } \rm{is}\mbox{ } \rm{selected} \\
0 & \rm{otherwise}.\\
\end{array}
\right.$$
Mathematically, we have to find an $n$-dimensional binary vector $x$ that maximizes $$\sum_{i=1}^n g_i x_i$$ among all binary vectors satisfying $$\sum_{i=1}^n w_i x_i \le c.$$ This problem is commonly referred to as the \emph{0-1 knapsack problem}. If it is possible to choose any fraction of an object, i.e. if the decision vector can be of the form $0\le x_i \le 1, \mbox{ } i=1,\ldots,n$, we call the problem a \emph{continuous knapsack problem}. The solution value of a continuous knapsack problem is clearly an upper bound for the solution value of the corresponding 0-1 knapsack problem.
\section{Maximizing the success probability under admissibility condition (Problem A)} \label{doe}

We will first present briefly how Problem A is treated using Neyman-Pearson lemmas in Föllmer and Schied \cite{Föllmer04}, pp.333-339. After that it will be shown that the same results can be accomplished through knapsack problem theory. Finally, we will discuss an approximative algorithm for solving Problem A. 
\subsection{Neyman-Pearson approach} \label{jek7}
The following result can be found with its proof in Föllmer and Schied \cite{Föllmer04}, p.335. Recall that $v$ is the upper bound for the initial payment we want to use to set up a hedging strategy.

\begin{theorem} \label{tpx8}
Assume that the set $\Gamma^A \in \mathcal{F}_T$ maximizes the probability $\mathbb{P}(\Gamma)$ among all sets $\Gamma \in \mathcal{F}_T$ satisfying the constraint
\begin{equation} \label{uc1}
 \mathbb{E}^*(H \cdot 1_{\Gamma}) \le v.
\end{equation}
Then the replicating strategy $\overline{\xi}^A$ of the option $H^A:=H \cdot 1_{\Gamma^A}$ solves Problem A. Moreover, $\Gamma^A=\{V_T^A\ge H\}$, where $V^A$ is the value process of the strategy $\overline{\xi}^A$.
\end{theorem}

In Föllmer and Schied \cite{Föllmer04} the authors define the measure $$d\mathbb{Q}:=\frac{H}{\mathbb{E}^*(H)}d\mathbb{P}^*$$ and consider the generalized density $d\mathbb{P}/d\mathbb{Q}$ that is received from the Lebesque decomposition. For our purposes it is enough to know that in our finite probability space this density takes the form
\begin{equation} \label{pricecost}
\frac{d \mathbb{P}}{d\mathbb{Q}}(\omega_i):=\left\lbrace 
\begin{array}{ll}
p_i/q_i & \rm{if} \quad \mathbb{Q}(\omega_i)\not = 0 \\
+\infty & \rm{if} \quad \mathbb{Q}(\omega_i)=0,\\
\end{array}
\right.
\end{equation}
where 
\begin{equation}\label{nr93}
p_i:=\mathbb{P}(\omega_i)
\end{equation}
and
\begin{equation}\label{nr95}
q_i:=\mathbb{Q}(\omega_i)=\frac{\mathbb{P}^*(\omega_i)H(\omega_i)}{\mathbb{E}^*(H)}.
\end{equation}
Once $\alpha:=v/\mathbb{E}^*(H)$ is defined, the level 
\begin{equation} \label{defn}
c^*:=\inf \left\{c \in \mathbb{R}_+|\mathbb{Q}\left(\frac{d\mathbb{P}}{d\mathbb{Q}}>c\cdot\mathbb{E}^*(H)\right)\le\alpha\right\}
\end{equation}
is introduced and it is shown, using Neyman-Pearson lemma, that if 
\begin{equation} \label{nwo5}
\mathbb{Q}(d\mathbb{P}/d\mathbb{Q}>c^*\cdot \mathbb{E}^*(H))=\alpha,\\
\end{equation}
then $\Gamma^A=\left\{d\mathbb{P}/d\mathbb{Q}>c^*\cdot \mathbb{E}^*(H)\right\}$ is an optimal set described in Theorem \ref{tpx8}.

To overcome the case when equation \eqref{nwo5} is not satisfied, the indicator function $1_{\Gamma^A}$ is replaced by a randomized test, i.e. an $\mathcal{F}_T$-measurable function $\psi$ such that $0\le\psi\le1$. After defining $\mathcal{R}$ as the set of all randomized tests, the authors consider the optimization problem of finding a randomized test $\psi^* \in \mathcal{R}$ that maximizes the expectation $\mathbb{E}(\psi)$ among all $\psi \in \mathcal{R}$ satisfying the constraint $\mathbb{E}^*(H \cdot \psi) \le v$. Such an optimal randomized test is by generalized Neyman-Pearson lemma given by
\begin{equation} \label{jep}
\psi^{NP}=1_{\{d\mathbb{P}/d\mathbb{Q}>c^*\cdot\mathbb{E}^*(H)\}}+\gamma \cdot 1_{\{d\mathbb{P}/d\mathbb{Q}=c^*\cdot\mathbb{E}^*(H)\} }
,
\end{equation}
where 
$$\gamma=\frac{\alpha-\mathbb{Q}(d\mathbb{P}/d\mathbb{Q}>c^*\cdot \mathbb{E}^*(H))}{\mathbb{Q}(d\mathbb{P}/d\mathbb{Q}=c^*\cdot \mathbb{E}^*(H))}.$$
Note that $\mathbb{Q}(d\mathbb{P}/d\mathbb{Q}=c^*\cdot \mathbb{E}^*(H))>0$ in our finite probability space, i.e. $\gamma$ is well-defined.

Finally it is shown that the replicating strategy $\overline{\xi}^*$ of the option $H^*:=H \cdot \psi^*$ actually maximizes the expectation of the so called success ratio $\psi_V$ among all admissible strategies with $V_0 \le v$ and that the success ratio $\psi_{V^*}$ of $\overline{\xi}^*$ coincides with $\psi^*$. 

\subsection{Knapsack approach}

In our finite market model the problem of finding an optimal set $\Gamma^A$ described in Theorem \ref{tpx8} is, in fact, a 0-1 knapsack problem. To see this, note that for any $\Gamma \in \mathcal{F}_T$ we have that $$H1_{\Gamma}=\sum_{\omega_i \in \Gamma}H1_{\omega_i}.$$ Thus, using \eqref{nr95} we can write the constraint \eqref{uc1} in form $$\sum_{\omega_i \in \Gamma}q_i \le \alpha:=\frac{v}{\mathbb{E}^*(H)}.$$  Further, $\mathbb{P}(\Gamma)$ can be written as $ \sum_{\omega_i \in \Gamma}p_i,$ where $p_i$ is as in \eqref{nr93}.

Since $\Omega$ consists of $n$ elements, we see that finding an optimal success set $\Gamma$ is equal to finding an optimal $n$-dimensional binary vector $x^A$. The problem of finding $\Gamma^A$ can thus be written as the following 0-1 knapsack problem:
\begin{problemap}
Find an $n$-dimensional binary vector $x^{A}$ that maximizes $$\sum_{i=1}^n p_i x_i$$ under the constraint $$\sum_{i=1}^n q_i x_i \le \alpha.$$
\end{problemap}
We will below use the notation $z^A$ for the value of the optimal solution, i.e.
\begin{equation} \label{gnu9}
z^A:=\sum_{i=1}^n p_i x_i^A=\mathbb{P}(V_T^A\ge H),
\end{equation}
where the latter equality follows from the fact $\Gamma^A=\{V_T^A \ge H\}$ mentioned in Theorem \ref{tpx8}.

Many numerical algorithms have been developed to solve 0-1 knapsack problems. A nice overview of some of these techniques can be found in Martello and Toth \cite{Martello} and in Martello et al. \cite{Martello00}. However, exact solution algorithms can be difficult to implement in a large probability space, where $n$ is huge. A good approximative algorithm will be given in Section \ref{greedy}.

\begin{remark}
It is often assumed (see e.g. Martello and Toth \cite{Martello}, p.14), that the variables $p_i$, $q_i$ and $\alpha$ in a 0-1 knapsack problem are positive integers. This assumption is a corner stone even for some numerical algorithms. However, in Problem $A'$ above we allow $0<p_i\le 1$ and $0\le q_i \le 1$. Algorithms that are based on the integer assumption, cannot naturally be used in this case. Note that our assumption $v<\mathbb{E}^*(H)$, or equivalently $\alpha<1$, rules out the possibility of the trivial solution $x_i=1$ for all $i=1,2,\ldots,n$.
\end{remark}

Next we will show that finding an optimal randomized test $\psi^*$ described in section \ref{jek7} can alternatively be seen as a continuous knapsack problem (cf. Section \ref{nqp}). The key point is that in our finite probability space there is a one-to-one correspondence between the set of all randomized tests and the set of all $n$-dimensional vectors $x$ satisfying $0\le x_i \le 1$ for $i=1,\ldots,n.$ This correspondence is simply given by equation 
\begin{equation} \label{nr96}
\psi(\omega_i)=x_i.
\end{equation}
Using \eqref{nr93} and \eqref{nr95} we see that the problem of finding an optimal $\psi^*$ is equivalent to finding an $n$-dimensional vector $x^*$ that maximizes $$\sum_{i=1}^n p_i x_i$$ under constraints $$\sum_{i=1}^n q_i x_i \le \alpha:=\frac{v}{\mathbb{E}^*(H)},\mbox{ } 0 \le x_i \le 1,\mbox{ } i=1,\ldots,n.$$

Assume now that the array $(\omega_1,\ldots,\omega_n)$ is ordered so that the quotient $p_i/q_i$ is non-increasing. Here we use the convention $p_i/q_i=+\infty$, if $q_i=0$. Thus, the most preferable states, the items that give the highest probability compared to cost, are placed first in the array. Suppose now that we consecutively choose the items, starting from the one giving the best probability over cost quotient and continuing until we find the first item $s$ that we no longer can afford to choose. In other words, we define the critical element
\begin{equation} \label{sbd}
s:=\min \left\{j:\sum_{i=1}^j q_i > \alpha\right\}.
\end{equation}
Due to assumption $\alpha<1$ we know that $1 \le s \le n.$ In Martello and Toth \cite{Martello}, pp.16,17 it is proved, by a simple contradiction, that an optimal solution to the continuous knapsack problem is given by
\begin{equation} \label{mfp1}
x_i^*=\left\lbrace 
\begin{array}{ll}
1,  & i=1,\ldots,s-1\\
\frac{\alpha-\sum_{j=1}^{s-1}q_j}{q_s}, & i=s\\
0, &  i=s+1,\ldots,n.\\
\end{array}
\right.
\end{equation}
We denote the optimal solution value by $$z^*:=\sum_{i=1}^n p_i x_i^*.$$ The value $z^*$ is an upper bound, the so called \emph{Dantzig's bound} for the optimal value $z^A$ in \eqref{gnu9}. If $\sum_{i=1}^{s-1}q_i=\alpha$, then $z^A=z^*$. Note the analogy with the Neyman-Pearson based result in \eqref{nwo5}.

Note that the difference between the optimal solutions $\psi^{NP}$ in \eqref{jep} and $\psi^*(\omega_i)=x^*_i$, $i=1,\ldots,n$ (cf. \eqref{mfp1}) is that the Neyman-Pearson approach tells us to search a critical level $L:=\{d\mathbb{P}/d\mathbb{Q}=c^*\cdot \mathbb{E}^*(H)\}$ whereas the knapsack approach suggests looking for a critical element $\omega_s$. The definitions of $c^*$ in \eqref{defn} and $s$ in \eqref{sbd} give that $\{\omega_s\} \subseteq L.$ In fact, there is no restriction on the values of an optimal randomized test $\psi$ on the critical set $L$ except that the level condition $$\mathbb{E}^{\mathbb{Q}}(\psi 1_{\{d\mathbb{P}/d\mathbb{Q}=c^*\cdot \mathbb{E}^*(H)\}})=\alpha-\mathbb{Q}(d\mathbb{P}/d\mathbb{Q}>c^*\cdot \mathbb{E}^*(H))$$
must be satisfied. This fact is mentioned also in Föllmer and Leukert \cite{Föllmer00}, p.126.

The particular form in \eqref{pricecost} that the derivative $d\mathbb{P}/d\mathbb{Q}$ takes in a finite probability space allows us to interpret the Neyman-Pearson result in compliance with the knapsack approach, i.e. that we choose to hedge against the states having the best probability over cost quotient. This ``cost-effectiveness'' interpretation of the Neyman-Pearson lemma is mentioned already in Kadane \cite{Kadane}.

\subsection{Greedy algorithm} \label{greedy}

We get an approximative solution to Problem $\rm A'$ if we set $x_s^*=0$ in \eqref{mfp1}. This technique is called the \emph{greedy algorithm} in Martello and Toth \cite{Martello}. The resulting solution clearly satisfies the cost constraint due to the definition of the critical element. Its solution value $$z^G=\sum_{i=1}^{s-1} p_i$$
satisfies $$z^G\le z^A \le z^*=z^G+\frac{\alpha-\sum_{i=1}^{s-1}q_i}{q_s}p_s\le z^G+p_s.$$ In other words, when using the greedy algorithm, the error is bounded above by $p_s$.

We can expect the greedy algorithm to work well in financial applications, since the probability space is usually relatively large, i.e. the probability for a single $\omega_i$ is small. For example, in a binomial model with $N$ steps, the maximal probability for an individual $\omega$ is $p_{max}=\max\{p,1-p\}^N$, where $p$ is the probability for an upward move during a single period. Even if $p$ would be as high as $0.9$, we would still have that, for instance for $N=100$, $p_{max}$ would be as small as $\approx 2.7 \times 10^{-5}$.

Note that we can in general reach higher probability by truncating the critical $x_s$ in \eqref{mfp1} than if we would truncate the entire set $L:=\{d\mathbb{P}/d\mathbb{Q}=c^*\cdot \mathbb{E}^*(H)\}$ in \eqref{jep}. Indeed, after choosing all the elements in $\{d\mathbb{P}/d\mathbb{Q}>c^*\cdot \mathbb{E}^*(H)\}$ we could still afford to choose some elements in $L$. 

However, from the computational point of view, it is often favorable to group together the elements having the same $p_i/q_i$-ratio. For example, consider a binomial model of $N$ steps. Then $\Omega$ consists of $2^N$ elements. If $H$ is an option whose value depends only on the value of the underlying asset at maturity, the ratio $p_i/q_i$ has the same value for all paths that lead to the same asset price at maturity. Thus, instead of ordering the $2^N$ elements separately, we may order the $p_i/q_i$-levels, whose number is at most $N+1$, and search the critical level $L$. Finally, we can study the elements in $L$ separately to see how many of them we can afford to hedge against. As a result, we obtain the same solution as with the greedy algorithm, but with less computational effort, since $N+1 \ll 2^N$.

\section{Maximizing the success probability when admissibility is not required (Problem B)} \label{lkjh}

Next theorem shows that we can with any initial capital $v_0 <\mathbb{E}^*(H)$ always create a so called \emph{quasi-replicating} strategy, in other words a self-financing strategy that replicates $H$ for all except one $\omega'$. This fact is then used to give a solution to Problem B. The notion quasi-replicating strategy is discussed in context of binomial model e.g. in Favero \cite{Favero01} and Favero and Vargiolu \cite{Favero06}.

\begin{theorem}
Take $v_0<\mathbb{E}^*(H)$ and an arbitrary $\omega' \in \Omega$. Then we can construct a strategy $\overline{\xi}$ whose value process $V$ satisfies $V_0=v_0$ and $V_T(\omega)=H(\omega)$ for all $\omega \in \Omega \setminus \omega'$. Moreover, $$V_T(\omega')=\frac{v_0-\mathbb{E}^*(H1_{\Omega \setminus \omega'})}{\mathbb{P}^*(\omega')}.$$
\end{theorem}
\begin{proof}
Note that $V_T(\omega')$ can be negative. However, due to completeness, there is a self-financing strategy $\overline{\xi}$ with value process $V$ that replicates the random variable $$H'=H1_{\Omega \setminus \omega'}+\frac{v_0-\mathbb{E}^*(H1_{\Omega \setminus \omega'})}{\mathbb{P}^*(\omega')} 1_{\omega'}.$$ (See Remark \ref{comp1}). Moreover, $V_0=\mathbb{E}^*(V_T)=\mathbb{E}^*(H')=v_0,$ since $V$ is a $\mathbb{P}^*$-martingale.
\end{proof}

\begin{corollary} \label{quasitheorem}
Let $v_0 \le v$ and $\omega_i$ such that $$\mathbb{P}(\omega_i)=\min_{\omega \in \Omega} \mathbb{P}(\omega).$$
Then the replicating strategy $\overline{\xi}^B$ for the random variable $$H^B=H1_{\Omega \setminus \omega_i}+\lambda 1_{\omega_i},$$ where $$\lambda=\frac{v_0-\mathbb{E}^*(H1_{\Omega \setminus \omega_i})}{\mathbb{P}^*(\omega_i)}$$ is a solution to Problem B.
\end{corollary}

\section{Minimizing expected shortfall under admissibility \\condition (Problem C)} \label{deo}

The following result is a simplified version of Theorem 8.10 in Föllmer and Schied \cite{Föllmer04}, p.341, where the result is proved for a general loss function and in a case where the market does not have to be complete.
\begin{theorem} \label{tpx8d}
Assume that there is a randomized test $\psi^C \in \mathcal{R}$ that maximizes the expectation $\mathbb{E}(H\psi)$ among all $\psi \in \mathcal{R}$ satisfying the constraint $\mathbb{E}^*(H \cdot \psi) \le v$. Then the replicating strategy $\overline{\xi}^C$ of the option $H^C:=H \cdot \psi^C$ solves Problem C.
\end{theorem}
An optimal randomized test can be found by using the generalized Neyman-Pearson lemma, as is done in Föllmer and Schied \cite{Föllmer04}, p.347. An alternative approach is to consider the problem of finding $\psi^C$ as a continuous knapsack problem. Recall \eqref{nr95} and \eqref{nr96} and define 
\begin{equation}\label{nr94}
m_i:=\mathbb{M}(\omega_i):=\frac{\mathbb{P}(\omega_i)H(\omega_i)}{\mathbb{E}(H)}.
\end{equation}
Then the problem of finding an optimal randomized test can be written in the following form.
\begin{problemcp}
Find an $n$-dimensional vector $x^C$ that maximizes $$\sum_{i=1}^n m_i x_i$$
under constraints $$\sum_{i=1}^n q_i x_i \le \alpha:=\frac{v}{\mathbb{E}^*(H)},\mbox{ } 0 \le x_i \le 1,\mbox{ } i=1,\ldots,n.$$
\end{problemcp}
This time the array $(\omega_1,\ldots,\omega_n)$ is ordered so that the quotient $m_i/q_i$ is non-increasing. Equations \eqref{nr95} and \eqref{nr94} give that $$\frac{m_i}{q_i}=\frac{\mathbb{E}^*(H)}{\mathbb{E}(H)}\frac{p_i}{p_i^*},$$ where $p_i=\mathbb{P}(\omega_i)$ and $p^*_i=\mathbb{P}^*(\omega_i)$. Thus, the states are ordered so that the quotient $p_i/p_i^*$ is non-increasing. The critical element $t$ is defined as $$t:=\min \left\{j:\sum_{i=1}^j q_i > \alpha\right\}$$ and an optimal solution is by Theorem 2.1 in Martello and Toth \cite{Martello}, p.16 given by 
$$x_i^C=\left\lbrace 
\begin{array}{ll}
1, & i=1,\ldots,t-1\\
\frac{\alpha-\sum_{j=1}^{t-1}q_j}{q_t}, & i=t\\
0, & i=t+1,\ldots,n.\\
\end{array}
\right.
$$
Note that $x^C$ does not in general coincide with $x^*$ in \eqref{mfp1} since the array $(\omega_1,\ldots,\omega_n)$ is ordered differently.
\section{Minimizing expected shortfall when admissibility is not required (Problem D)} \label{hjkl}

Favero \cite{Favero01} and Favero and Vargiolu \cite{Favero06} study the problem of minimizing expected shortfall in the special case of a binomial model, when admissibility is not required. Scagnellato and Vargiolu \cite{Scagnellato} discuss the same problem in a more general multinomial model. In those papers, the authors prove their results via dynamic programming. In this paper we provide an alternative approach by showing that even this problem can be reduced to a knapsack problem. To begin with, we state the following theorem.

\begin{theorem} \label{br5}
Assume that the random variable $X^D$ maximizes $\mathbb{E}(X)$ among all random variables $X$ that satisfy $X \le H$ and $\mathbb{E}^*(X)\le v.$ Then the replicating strategy $\overline{\xi}^D$ for $X^D$ solves Problem D.
\end{theorem}
\begin{proof}
Take any self-financing strategy $\overline{\xi}$ with value process $V$ such that $V_0 \le v$ and $V_T\le H$. The random variable $V_T$ satisfies $\mathbb{E}^*(V_T)=V_0\le v$ by the martingale property. Thus, we have by assumption that $$\mathbb{E}(V_T)\le \mathbb{E}(X^D).$$
On the other hand, for the strategy $\overline{\xi}^D$ with the value process $V^D$ we have $$V_T^D=X^D \le H$$ and $$V_0^D=\mathbb{E}^*(V_T^D)=\mathbb{E}^*(X^D)\le v.$$ Further, the maximal expectation is attained by using strategy $\overline{\xi}^D$, since $$\mathbb{E}(V_T^D)=\mathbb{E}(X^D).$$
\end{proof}

\begin{remark}
Note that we cannot use in Theorem \ref{br5} a similar approach via randomized tests that is used in Theorem \ref{tpx8d}. In Problem C the optimal solution is searched among strategies whose value processes $V$ satisfy $0 \le V_T \le H$, which enables us to express $V_T$ as the product of the claim $H$ and a randomized test $\psi$. In Problem D, however, the value process $V_T$ may become negative. On the other hand, the more general approach in Theorem \ref{br5} could be used to reduce Problem C. Instead of searching for an optimal randomized test we could look for an optimal random variable $X^C$ that maximizes $\mathbb{E}(X)$ among all random variables $X$ satisfying the constraints $0\le X \le H$ and $\mathbb{E}^*(X)\le v$. In other words, the only difference would be the additional constraint $X\ge 0$, which is connected with the admissibility condition.
\end{remark}

Denote $p_i:=\mathbb{P}(\omega_i)$, $p_i^*:=\mathbb{P}^*(\omega_i)$ and $h_i:=H(\omega_i)$. The problem of finding $X^D$ obviously takes the following form in our finite probability space: 
\begin{problemdp}
Find a vector $x^D$ that maximizes $$\sum_{i=1}^n x_i p_i$$
under constraints $$\sum_{i=1}^n x_i p_i^*\le v, \quad x_i \le h_i, \quad i=1,\ldots,n.$$
\end{problemdp}
This is almost similar to the continuous knapsack problem $\rm C'$. The only difference is that instead of constraint $0\le x_i \le 1$ we have $x_i \le h_i$, i.e. the values of the decision variables $x_i$ are unbounded below and bounded by a deterministic, but varying non-negative boundary above. The problem can be solved in a way that resembles the solution method for a continuous knapsack problem.

Assume that the array of the states $\omega_i$, $i=1,\ldots,n$ is ordered so that $$\frac{p_1}{p_1^*}\ge \ldots \ge \frac{p_n}{p_n^*}.$$ The following theorem gives a solution to Problem $\rm D'$.

\begin{theorem} \label{sol65b}
An optimal solution $x^D$ to Problem $\rm D'$ is given by
\begin{align} 
x^D_i&= h_i \quad for \quad i=1,\ldots,n-1 \nonumber \\
x^D_n&=\frac{v-\sum_{i=1}^{n-1}h_i p_i^*}{p_n^*}.\label{uj3b}
\end{align}
\end{theorem}
\begin{proof}
We prove our result in a way that resembles the proof of Theorem 2.1 in Martello and Toth \cite{Martello}. Firstly note that for any optimal solution $x$ it has to hold that 
\begin{equation} \label{hm3b}
\sum_{i=1}^n x_i p_i^*=v.
\end{equation}
Without any loss of generality we can assume that $p_i/p_i^* > p_{i+1}/p_{i+1}^*$ for all $i$. Let $x^*$ be an optimal solution to Problem $\rm D'$ and suppose that $x_k^*<h_k$ for some $k<n$. Now if we take $\epsilon >0$ small enough, we could increase $x_k^*$ by $\epsilon$ and decrease $x_n^*$ by $\epsilon p_k^*/p_n^*$. But this would increase the value of our objective function by $\epsilon(p_k-p_n p_k^*/p_n^*)$ ($>0$ since $p_k/p_k^* > p_n/p_n^*$) and give us a contradiction. Therefore, $x_k^*=h_k$ for $k<n$ is necessary for an optimal solution $x^*$. The statement \eqref{uj3b} follows from \eqref{hm3b}.
\end{proof}

Thus, we have proved that if we choose $\omega_i$ such that $$\frac{d\mathbb{P}}{d\mathbb{P}^*}(\omega_i)=\min_{\omega \in \Omega} \frac{d\mathbb{P}}{d\mathbb{P}^*}(\omega),$$ then the optimal strategy solving Problem D is by Theorem \ref{br5} the replicating strategy $\overline{\xi}^D$ for the random variable $$H^D=H1_{\Omega \setminus \omega_i}+\varphi 1_{\omega_i},$$ where $$\varphi=\frac{v-\mathbb{E}^*(H1_{\Omega \setminus \omega_i})}{\mathbb{P}^*(\omega_i)}.$$
Note that this result coincides with the result in Scagnellato and Vargiolu \cite{Scagnellato}, pp.148,149, where the authors prove it for a complete multinomial model via dynamic programming.

The knapsack problem approach helps us to see the difference between the optimal strategies when admissibility is or is not required, i.e. the difference between the solutions to Problems C and D. In both cases we arrange the states similarly, such that the ratio $\mathbb{P}/\mathbb{P^*}$ is non-increasing. Then we choose consecutively $\omega$'s for which we want to hedge the option perfectly, starting from $\omega_1$ that has the largest ratio. When admissibility is required, the critical element $\omega_t$ is the first one that we no longer can afford to choose. The remaining capital is then used to construct a partial hedge for this state. If the admissibility condition is dropped, we can continue choosing $\omega$'s until we reach the last one, $\omega_n$, having the least ratio. The value $H^D(\omega_n)$ is then adjusted so that the cost constraint is satisfied.

\section*{Acknowledgements}
I want to thank my supervisor Paavo Salminen for guidance and stimulating discussions and Esko Valkeila for bringing the idea of quasi-replicating strategies to my attention.

\end{document}